\documentclass{emulateapj}
\usepackage{graphicx}

\newcommand{\fig}[1]{Figure ~\ref{fig:#1}}
\shorttitle{The Archival Pure Parallels Program}
\shortauthors{Wadadekar et. al}
\begin{document}

\title{The WFPC2 Archival Parallels Project}
\author{Yogesh Wadadekar, Stefano Casertano, Richard Hook\altaffilmark{1}, B\"ulent K{\i}z{\i}ltan\altaffilmark{2}, Anton Koekemoer, Harry Ferguson and Doichin Denchev}
\affil{Space Telescope Science Institute, 3700 San Martin Drive, Baltimore, MD 21218}
\altaffiltext{1}{Space Telescope European Coordinating Facility, European Southern Observatory, Karl Schwarzschild Str-2, D-85748 Garching, Germany}
\altaffiltext{2}{Department of Astronomy and Astrophysics, University of California, Santa Cruz}

\begin{abstract}

We describe the methods and procedures developed to obtain a
near-automatic combination of WFPC2 images obtained as part of the WFPC2
Archival Pure Parallels program.  Several techniques have been developed or
refined to ensure proper alignment, registration, and combination of
overlapping images that can be obtained at different times and with
different orientations.  We quantify the success rate and the accuracy
of the registration of images of different types, and we develop techniques 
suitable to equalize the sky background without unduly affecting
extended emission.  About 600 combined images of the 1,500 eventually
planned have already been publicly released through the STScI Archive. 
The images released to date are especially suited to study star
formation in the Magellanic Clouds, the stellar population in the halo
of nearby galaxies, and the properties of star-forming galaxies at $ z
\sim 3 $. 

\end{abstract}
  

\section{Introduction}

\subsection{The WFPC2 Archival Pure Parallels Program}

Over a span of several years, from 1997 through 2003, the Wide Field
and Planetary Camera 2 (WFPC2) on board the Hubble Space Telescope
(HST) carried out an Archival Parallel Program under the auspices of
the Parallels Working Group chaired by Jay Frogel.  The Archival
Parallel Program consisted of a large number of parallel images, i.e.,
images of the area in the sky towards which the camera was pointed
while another instrument on HST was executing planned observations.
Such pointings are constrained to be random, in the sense that they
are not expected to contain any special sources-- except by reason of
proximity to the primary target-- which is 5 to 12 arcmin away
depending on the instrument used.

The program was designed to provide a set of observations that would
provide a valuable data base for the HST archive with the potential to
impact a range of scientific programs that the community at large could
carry out\footnote{The final report of the Working Group is available at {\tt
http://www.stsci.edu/instruments/parallels/HSTParallel.html };
additional material is available at {\tt
http://www.stsci.edu/instruments/parallels/ }.}.  The Parallel Working
Group identified three areas of special interest: young stars and star
forming regions in our Galaxy and other nearby galaxies; the stellar
content of galaxies  in the local Universe (including our own); and large
scale structure in the universe and the distribution and evolution of
galaxies.

For WFPC2, the observations recommended by the Parallels Working Group
were implemented in three different programs: Galactic, extragalactic,
and special objects.  The Galactic and extragalactic programs were
used for generic pointings, i.e., those not in the special objects
category; Galactic pointings are those at Galactic latitude $ | b | <
20\deg $, extragalactic those at higher latitudes.  Special objects
pointings are those that fall close to objects of interest; the most
common category for the WFPC2 parallels is that of pointings within a
specified distance--- 10 arcmin, except for a few of very large
galaxies such as M31--- of galaxies less than about 3 Mpc away.

The Parallel Working Group specified the observing strategy for each
type of pointing.  In general, observations were obtained in one or more
of the four Hubble Deep Field filters, F300W, F450W, F606W, and F814W,
depending on the available time-- i.e., the length of the primary
observation; other filters were used for some of the programs.  In
almost all cases, the emphasis was on breadth of the survey, taking
advantage of the expected large area coverage, rather than depth, which
could not match that of dedicated observations.  Two features were
common to all programs: exposures were always obtained in pairs for each
filter, in order to facilitate data processing and especially rejection
of cosmic rays; and regardless of the program, the most sensitive
filter, F606W, was always used first, although possibly for brief (300s)
exposures.

Thereafter, the observing strategy varied depending on the type of
pointing.  For extragalactic pointings, most observations were obtained
as part of the Broad Band Survey, which aimed at obtaining multicolor
data in the four Hubble Deep Field filters for relatively bright ($ V
\sim 24 $ mag), and therefore uncommon, galaxies.  The four filters   
contribute depth (F606W), UV morphology and U dropouts (F300W), B   
morphology and dropouts (F450W), and photometric redshift information
(F814W), and were generally obtained in that order of priority.  A small
fraction of observations were obtained in a special program targeting
the medium-width filters F410M and F467M, which could identify
star-forming galaxies at redshift 2.36 and 2.85, respectively.

\begin{figure*}
\plotone{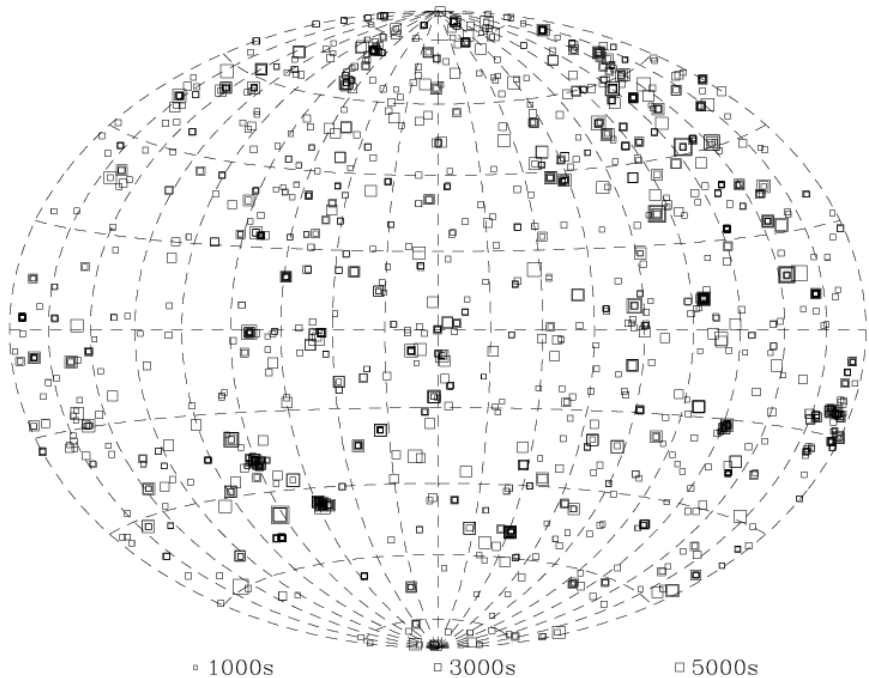} 
\caption{Distribution of APPP datasets on the sky in galactic coordinates. The size of each point is scaled by the total exposure time at that location. The two concentrations in the bottom left quadrant are the Large and Small Magellanic Clouds.}\label{fig:skydistribution}
\end{figure*}

For Galactic pointings, a selection of narrow-band filters
(F502N, F656N) was added to the extragalactic filters in order to study
the properties of diffuse line emission in objects such as distant
star-forming regions and planetary nebulae, which could expect to be
observed relatively often in random fields at low Galactic latitude.

The special pointings in the vicinity of bright, nearby galaxies were
chosen so that WFPC2 could resolve the stellar population into
individual stars, thus permitting a study of their resolved stellar
populations.  For Local Group galaxies, the sensitivity achieved in a
typical 1000 sec observation is sufficient to detect young stars in
H-$\alpha$ emission, and thus F656N was added to the complement of
filters.  In a few cases, this has resulted in spectacular
observations of diffuse line emission in the Magellanic Clouds.  For
galaxies beyond the Local Group, the sensitivity is insufficient to
take full advantage of F656N or of F300W, and thus only F606W, F450W,
and F814W were used.

As of July 2002, the WFPC2 archival parallels
had accumulated over 14,000 separate exposures with a total exposure
time of 11.77 Msec.  Of this, the primary broadband filters F300W,
F450W, F606W, and F814W had exposure times of 2.39, 1.42, 6.67 and 0.65
Msec respectively.  The distribution of these observations over the sky
is shown in \fig{skydistribution}. 

\subsection{The Archival Pure Parallels Project}

However, the enormous scientific potential of the archival parallels
has remained largely untapped. The primary reason is accessibility: the
WFPC2 images are generally not available in a readily usable form. The WFPC2
images in the archive need reliable rejection of image artifacts such
as cosmic rays and hot pixels; the combination of co-aligned and
non-coaligned exposures, and source catalog information. Some of
these services are available for a subset of WFPC2 observations from
the WFPC2 Associations effort\footnote{{\tt http://cadcwww.dao.nrc.ca/wfpc2/WFPC2\_pipe.html}}.

The Archival Pure Parallels Project (hereafter APPP) is an ongoing HST
Archival Legacy program that aims at processing, combining, and
delivering a large fraction of the parallel images taken by the WFPC2
Archival Parallels Program.  The Project will prioritize the available
pointings on the basis of number of filters available, length of
integration, and pointing characteristics.  Among the special classes
of observations that will receive a high priority are fields in the
Magellanic Clouds, which will permit a much wider study of star
formation in different regions of the Clouds; fields close to known
radio and X-ray sources ---about 40 sources in the FIRST catalog fall
into areas observed for the WFPC2 Archival Parallels--- fields that
overlap with Chandra and XMM observations; and fields that have data
in more than two filters\footnote{An overview of the goals of the
Project can be found at {\tt
http://www-int.stsci.edu/$\sim$yogesh/APPP/}}.  Overall, the Project
will make available to the community a total of 2500 images in 1500
pointings, encompassing about 7000 separate exposures, or about half
of the Archival Parallels images.  About 600 images in 150 pointings
have been released to date through the Multimission Archive at the
Space Telescope (MAST); the images can be obtained through the APPP
page\footnote{{\tt http://archive.stsci.edu/prepds/appp/}} at the
MAST. We expect that about 3 years (FTE) of labor will be expended for
the completion of this project.

The APPP is in many ways complementary to the WFPC2 associations
effort (Micol et al. 2000).  Its scope is more limited, since it
specializes to images taken as part of the Archival Pure Parallels
programs, while the WFPC2 associations project extends to all WFPC2
images.  On the other hand, the APPP images are in a single frame,
while those produced by the WFPC2 associations effort are separated by
chip; the APPP images are corrected for geometric distortion; and the
APPP combines images taken as part of different visits, even when
taken at different orientations.  For regions of the sky observed
repeatedly, this approach results in deeper delivered images that can
also cover seamlessly a larger area of the sky; cross-registration
between filters also allows source colors to be measured directly even
for non-point sources.  A more detailed comparison of the final products of the WFPC2 associations and the APPP is presented in Section~4.

In this paper we describe in detail the methodology and procedures
used to produce the combined images that have been released and that
will be used for all upcoming images.  In Section~2, we describe the
techniques that we have developed for accurate image registration,
background equalization and cosmic ray rejection.  In Section~3, we
explain the data processing steps in procedural form.  Section~4
summarizes the quality control procedures that we use, before the data
is publicly released.

\section{Data reduction processes}

The aim of the data processing pipeline is to produce astrometrically
registered, drizzled images with background equalization and reliable
rejection of artifacts such as cosmic rays and hot pixels, for all of
the four WFPC2 chips. Our techniques have been optimized to exploit the
strengths of the WFPC2 instrument while simultaneously trying to
mitigate its deficiencies.

\subsection{Alignment across chips}

It is known from from science data of rich stellar fields as well as
from Kelsall spot data that the four WFPC2 detectors move with respect
to each other over time, presumably as a consequence of physical
changes in the WFPC2 optical bench. Casertano \& Wiggs (2001) used the
positions of 43 Kelsall spots in 173 images taken throughout the life
of WFPC2, approximately one every two weeks. They found that the shift
in relative positions of the spots (and therefore the chips) with time could be
effective modeled using a fourth degree polynomial.

We use the coefficients for the fourth degree polynomial from
Casertano \& Wiggs (2001) to determine the relative position of the
chips as the time of observation. 

\subsection{Alignment of images}\label{sect:alignment}

The nominal World Coordinate System (WCS) values for the WFPC2 are subject to small errors of
typically 0.1 arcsec or smaller. In a few instances, where different
guide stars are used to obtain WCS coordinates for two overlapping
images, the relative errors in position may be as high as 2-3
arcsec. We have developed a procedure to correct for these astrometric
errors that is described below.

The image registration is best performed using individual images
wherein all four chips have been drizzled to the output frame. We first
construct a sequence of images such that each successive image in the
sequence has the maximum possible overlap with one or more of the
preceding images in the sequence. Thus if there are $n$ datasets, a set
of $(n-1)$ pairs with optimal overlap are determined. Relative shifts
and rotations between images are determined pairwise and then
propagated as described below. The motivation for choosing a maximum
overlap sequence is to maximize the number of matched sources during
image registration.

We have developed and tested two distinct approaches to image registration:

Approach 1: We first determine the centroid position for all sources
in the reference image and the image to be shifted using  SExtractor 
(Bertin \& Arnouts 1996). During source extraction, it is critical to
use the output weight data from drizzle as the {\tt WEIGHT\_IMAGE} in
 SExtractor, to reduce the number of spurious source detections. We
then filter the source lists to exclude sources with possibly
incorrect positions (e.g. saturated sources) and pass the source
centroid positions to a triangle matching program that attempts to
find matching sources in the reference and shifted images. This
algorithm uses the principle that similarity properties of triangles
hold for any transformation that involves a shift and/or a
rotation. Once a matched source list is obtained, this list is passed
to the IRAF {\it geomap} program with {\tt fitgeometry=rotate}, which
determines the relative shift and offset between the two images. This
approach works well when the number of real sources detected in the
image is comparable to (or exceeds) the number of cosmic rays. Such a
situation exists only for the minority of low galactic latitude and
Magellanic cloud pointings in our data. Pointings, where there are few
real sources and many more cosmic rays, are predominant. In such
situations, the triangle matching method fails to find a sufficient
number of matched triangles and sources for nearly 50\% of
pointing/filter combinations. We therefore explored an alternative
approach to image registration.

\begin{figure*}
\plotone{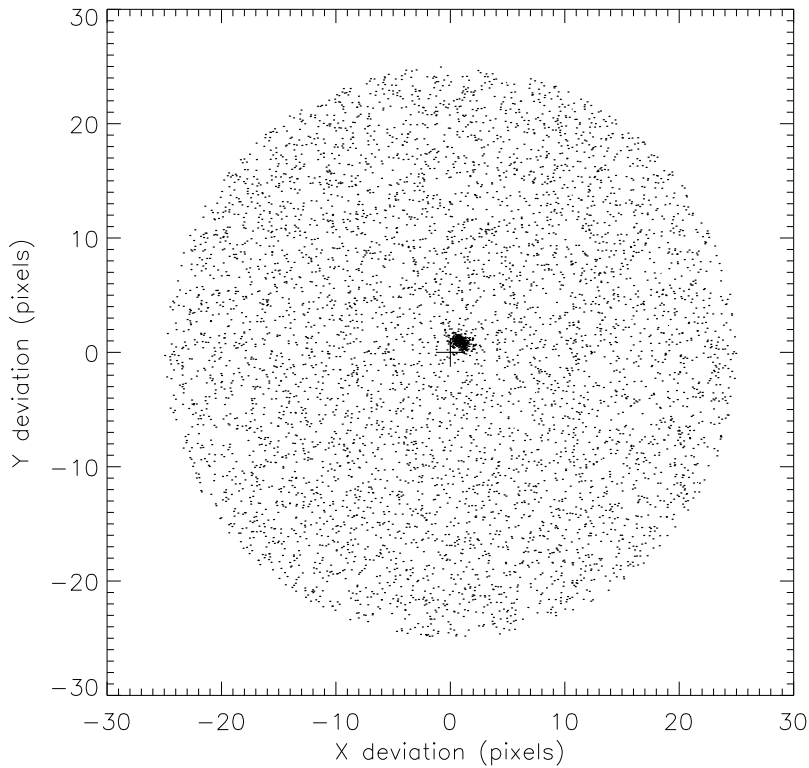} 
\caption{Deviations of sources within 25 arcsec of sources in the reference image. The cluster of points near the center represents true source matches between reference image and the source image. Most other points are chance coincidences of cosmic rays. }\label{fig:deviations}
\end{figure*}

Approach 2: Centroid positions for sources in the reference and the
shifted image are obtained and filtered as in Approach 1. For each
source in the reference image, we search for sources that lie within
25 pixels (2.5 arcsec) in the shifted image. These sources are plotted
as deviations in $x$ and $y$ coordinates with respect to the source in
the reference image. A sample deviation plot is shown in
\fig{deviations}. We expect that the distribution of real sources in
this deviation plot will cluster, while the cosmic rays will tend to
be distributed randomly. An inherent assumption in using this
approach, is that relative rotations between images are small. This is
a valid assumption because the orientation angle of the telescope
(ORIENT) is known to a precision of 0.1 degrees or better. Note that
we still fit for a shift and a rotation, although the rotation is
quite small.  We construct a 2-D histogram of deviations with a
binsize of 1x1 pixel and find the peak. If no unique peak exists, or
there are too few points in the peak, we increase the binsize to 2x2
pixels (and 3x3 pixels if required). We find that over 95\% image
registrations are successfully obtained with a binsize of 1x1
pixel. Once the appropriate peak is identified, all sources with
deviations lying within the peak are designated matched sources and
supplied to {\it geomap}, again with {\tt fitgeometry=rotate}. {\it
geomap} is set to reject outliers in its fit, with {\tt maxiter=5} and
{\tt reject=2.0}. This approach is successful in registering $>95$\%
images in F450W or redder broadband filters, and about 75\% of images
in the F300W filter. The somewhat lower success rate in registering
images in the F300W filter is primarily due to the dramatically lower
throughput of WPFC2 in this filter, relative to redder
filters. Overall, there is a dramatic increase in registration success
rate as compared to the triangle matching approach. We therefore
settled on using this image registration approach in our pipeline.

The image registration is then repeated pairwise e.g. Given a set of say 4
images, the two images with the highest overlap, say Images A and B
are identified and Image B is registered relative to Image A. Then,
amongst the remaining 2 images the image with the highest overlap with
either A or B is chosen e.g. we may find that Image D has the highest
overlap with Image A. Then Image D is registered to Image A. The
remaining Image C is registered to one of Images A, B, D with which it
has the highest overlap. Pairwise shifts and rotations obtained from
{\it geomap} are adjusted such that they are relative to the one image that
does not undergo a shift or rotation (Image A in our example). These
shifts and rotations are transformed to a corresponding change in the
WCS of the 4 chips. In the above example, all chips of all images
except those of Image A, would receive revised WCS values.

\subsection{Sky Background Equalization}\label{sect:backequal}

The observed sky background at a particular location on the sky can
change due to light scattered into the aperture from the bright earth
limb. Due to this phenomenon, two images of the same part of the sky
taken at different times may have different values of the sky
background. It is important to correct for such differences to ensure
a uniform sky in the drizzled output image and to accurately reject
cosmic rays.

We correct for relative offsets between sky values by matching
the sky pairwise in the maximum overlap sequence as obtained
above. The sky offsets between images is the median of the difference
between valid pixels in the two images in the region of overlap. Sky
offsets are determined relative to the overall reference image (Image
A in our example) and corrected for.

\begin{figure*}
\plotone{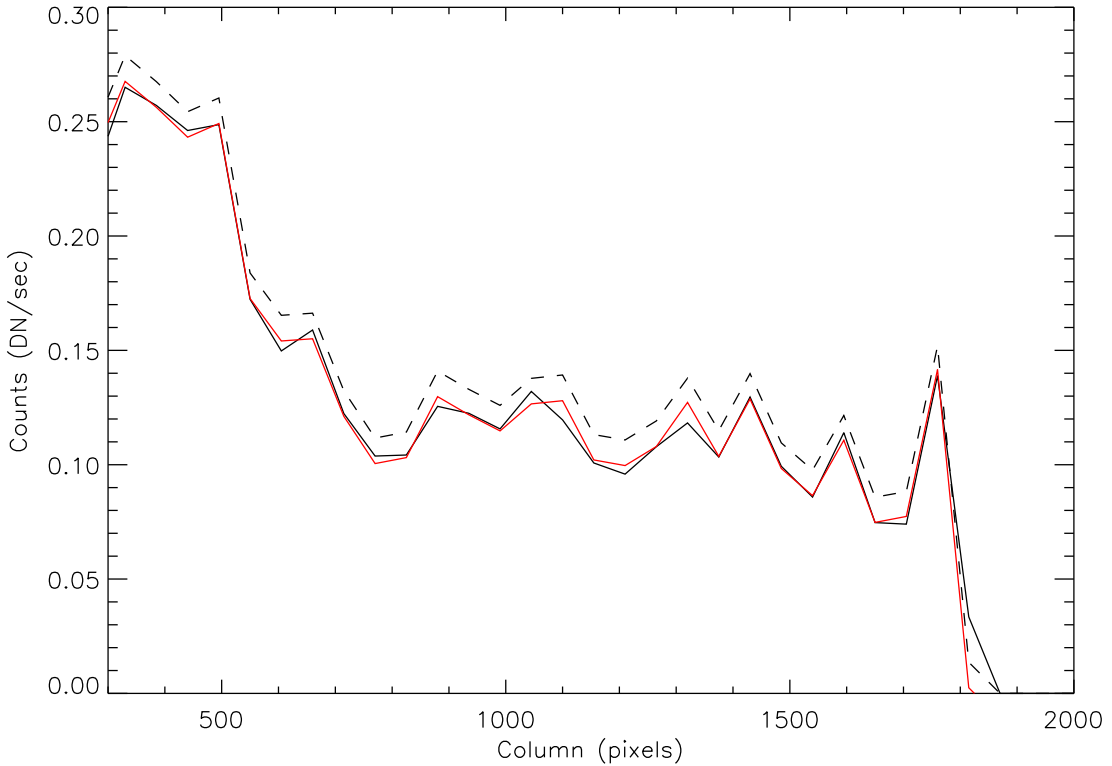} 
\caption{Plot of counts (DN/sec) for a single row of image data. The data have been binned for clarity. The black solid line depicts the source counts in the reference image. The dashed line shows the counts in the shift image which are all systematically higher. The red solid line depicts counts in the shift image after the sky background equalization process.}\label{fig:backequal}
\end{figure*}

We illustrate the background equalization process in
\fig{backequal}. We plot the counts in a particular row of a reference
image (black solid line) and the counts in the same row of the
comparison image (dashed line) that has a higher value of the sky
background. The higher average counts at lower column values is due to
extended nebular emission in that part of the image. After the sky
background offset has been corrected for, the counts in the comparison
image substantially match those in the reference image (red solid
line). The figure only illustrates this process for one row of binned
data, but in the pipeline, this procedure is carried out using all
pixels in the region of overlap.

Note that only relative sky offsets are corrected by our procedure. No
attempt is made at subtracting the sky background from the reference image.

\subsection{Drizzling}

Drizzle is a method for the linear reconstruction of an image from
undersampled, dithered data (Fruchter \& Hook 2002). The algorithm,
known as Variable-Pixel Linear Reconstruction, or informally as
``Drizzle,'' preserves photometry and resolution, can weight input
images according to the statistical significance of each pixel, and
removes the effects of geometric distortion on both image shape and
photometry. In addition it provides a method for robust cosmic ray
rejection. Drizzle was first used in the analysis of the Hubble Deep
Field North data (Williams et. al 1996). It has since been included as
a part of the STSDAS software package.

Drizzle takes the input data and transfers each pixel to the output
frame. This typically involves a shift (in x and y direction), a
rotation and a scaling. In addition, each input data pixel may be shrunk
before it is drizzled. During drizzling, the user needs to specify the
shifts, rotations and scaling needed to go from the input frame to the
output frame. A version of drizzle that simplifies this process by
working directly with WCS parameters has recently been developed.
This code is called {\it wdrizzle}. While using {\it wdrizzle}, the
user needs to only specify the WCS keywords of the input frame
(available in the image header) and the WCS keywords of the output
frame.

We ran {\it wdrizzle} on all 4 chips of each input dataset to its own
output image (as determined by the output WCS) and weighted by the
inverse variance map. A {\tt scale} and {\tt pixfrac} value of 1.0 was
used throughout, along with a square kernel. Kozhurina-Platais et
al. (2003) used the inner calibration field of $\omega$ Cen exposed
through filters F300W, F555W and F814W to determine the geometric
distortion of WFPC2 as a function of wavelength. They incorporated the
improved PSF-fitting technique of Anderson and King (2002) to fit a
bicubic polynomial model to derive geometric distortion coefficients
in the F300W and F814W filters relative to the distortion-free
coordinates in the F555W filter. We supply these distortion
coefficients to {\it wdrizzle}. If coefficients are unavailable for a
particular filter, those from the nearest available filter are used.

\subsection{Cosmic ray rejection}\label{sect:crrej}

We adopt the procedure proposed by Fruchter \& Hook (2002) for cosmic
ray rejection.

We construct the median image by performing the median operation on
the sky offset corrected images. The medianmask allows us to exclude
invalid pixels from the median image. The median image is relatively
free of cosmic rays. We back propagate the median image to the input
frame of each chip of each of the individual input images, taking into
account the image shifts/rotation and geometric distortion. This is
done by interpolating the values of the median image using a program
called {\it wblot} (which is a WCS aware version of the {\it blot}
program in the dither package).

We take the spatial derivative of each of the blotted images. The
derivative image is useful for estimating the extent to which errors in
the computed image shift or the blurring effect of the median
operation, have modified the counts in the blotted image. We compare
counts in each original image with those in the corresponding blotted
image. If the difference is larger than that expected by a combination
of the expected noise, the blurring effect of taking the median, or an
error in the shift, the pixel is flagged as cosmic-ray affected. We
repeat this step on pixels adjacent to pixels already flagged, using a
more stringent comparison criterion.

\section{Data reduction pipeline}

All the techniques described above were implemented as a fully
automated data reduction pipeline which can be applied to the
calibrated data obtained from the STScI Archive.  The pipeline
consists of two segments: a database management unit, developed in
Python, which identifies datasets that are proximate on the sky and
arranges the appropriate data location and a data processing unit,
written primarily in IDL, which performs the data alignment and
combination.  The latter uses callable versions of the {\it wdrizzle}
and {\it wblot} programs, identical in function to the STSDAS 
versions.  Most of the pipeline processing is straightforward and
sequential; however, as described above in
Section~\ref{sect:alignment}, the image registration step is fragile,
and can fail for a variety of reasons (poor initial positions, lack of
sources, insufficient exposure time, insufficient image overlap).
Automatic checks verify whether registration was successful; if not,
the affected images---in some cases a full pointing/filter
combination---are excluded from further processing.  With current
procedures, about 10\% of all images fail the automatic registration
check. The majority of registration failures are in the F300W
filter. An additional 5\% fail the subsequent manual quality
check (see Section~4). There are no failures in any of the other steps. 
While future changes in our procedures could
slightly improve our success rate, our experience to date suggests that
a fully automated alignment procedure is unlikely to work under all
possible circumstances.

We now proceed to describe the individual steps taken as part of the 
image processing pipeline.

\begin{figure*}
\plotone{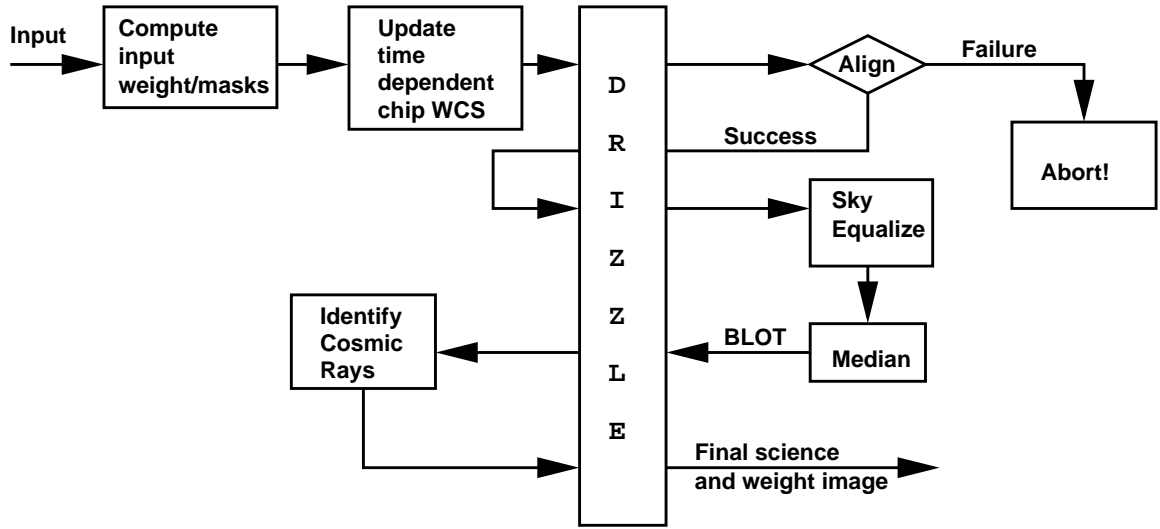} 
\caption{Data flow diagram for the major steps in the Stage 1 pipeline processing. There are three separate drizzle operations and one inverse drizzle operation (blot). Each step is described in detail in the text.}\label{fig:pipeline}
\end{figure*}

\subsection{Image database and pointings}

The first step in the processing is the definition of the relevant
pointings and of the images associated with each.  We started by
extracting from the MAST archive\footnote{\tt
http://archive.stsci.edu} the data files associated with all images
obtained as part of the Archival Pure Parallels, including both the
data files (suffix {\tt \_c0f.fits}) and the data quality files
(suffix {\tt \_c1f.fits}).  We also obtained all the flat field files
used in processing these images. All images were retrieved from MAST
during a brief period in July 2003. This ensured that the on-the-fly
reprocessing (OTFR) performed automatically by the MAST archive was
identical for all datasets.

After retrieving all images, we extracted all header parameters,
including both primary and group parameters, and captured them into a
relational database using MySQL.  The datasets were then partitioned
into groups using a simple proximity relation.  First, two datasets
are related if their reference point, the position of the WFALL
reference point on the sky, is less than $ 80 \arcsec $ apart.
Second, two datasets related to a third are considered related.  This
proximity relation is symmetric and transitive, and therefore it is an
equivalence relation---thus defining a partition in the set of all
images under consideration.  Each of these equivalence groups is
called a {\it pointing}.  In practical terms, pointings are defined by
starting with datasets less than $ 80 \arcsec $ apart, and then
extending the rule by the friend-of-friends algorithm i.e a dataset
belongs to a pointing if and only if it is less that $ 80 \arcsec $
apart from at least one other dataset in that pointing.

For the purpose of pointing definition, all datasets are considered
together, regardless of the filter used or the length of exposure. This
could result in some anomalies, where combined images in each filters
are disjoined, non-overlapping, or could contain only one dataset.  We
find that such anomalies are exceedingly rare, and thus the definition
of pointing we adopt is useful.  Each pointing is assigned a unique
9-character ID based on the name of the dataset in that pointing with
the earliest observation date. 

With our criteria, the 14965 datasets in our database are grouped into
2460 pointings, with an average of about 6 datasets per pointing. The
largest pointing contains 161 datasets and spans a diameter of 12
arcmin on the sky.

Of these, 2305 pointings have at least one filter with two or more
datasets each longer than 100s in that filter. All these pointings may
be processed by our pipeline. However, we are prioritizing the
processing based on the science drivers, total exposure time and the
number of filters available. A total of 573 images in 149 pointings
have already been released to the MAST archive. We anticipate that a
total of about 2500 images in 1500 pointings will be released before
the completion of this project.

\subsection{Preparing data for pipeline processing}

Once a pointing is chosen for pipeline processing, several steps
need to be undertaken in order to simplify the processing
itself and to obtain uniform output products.  These steps
consist of collecting together the relevant data, defining the
output images, and preprocessing the data to the extent needed.
We have developed a number of Python scripts to carry out these
steps automatically and efficiently.

Data collection is directory-based.  Each pointing is assigned
a directory, with a separate subdirectory for each filter.
The subdirectory contains the relevant science, data quality,
and inverse flat field images in IRAF group format (suffixes 
{\tt c0h}, {\tt c1h}, and {\tt r4h} respectively).  

In each subdirectory, subsets of images which are nominally coaligned
(shifts of less than $0.01\deg $ in roll angle and $ 0.01\arcsec $ in
position) are considered suitable for CR-rejection, which is then
performed using the standard STSDAS {\it crrej} task.  Using the
coaligned cosmic ray rejection reduces significantly the number of
images that fail the alignment step, and therefore increases
significantly the quality of the resulting combined images.

Finally, the output image needs to be defined.  The output image is
rectangular, with pixel size equal to the average pixel size in chip
3, and is oriented with North up and East to the left.  The output
image for each filter is large enough to accommodate all of the input
images, with sufficient margin to account for possible alignment
refinements (see Section~\ref{sect:alignment}). Output image sizes
range from 1600 to 7200 pixels on a side.

\subsection{Stage 1 pipeline}

The main steps of the Stage 1 pipeline are summarized as a data flow
diagram in \fig{pipeline}. The Stage 1 pipeline includes three separate drizzle operations and one inverse drizzle operation (blot). The various steps of the procedure are described below.
 
\subsubsection{Bad pixels and variance maps}

Bad pixels in an image are those that satisfy at least one of the
following conditions:

 \begin{enumerate}
\item have been flagged as bad in the data quality files.  These
include transmission and other failures, blocked columns, saturated
pixels and bad pixels listed in calibration reference files. 
\item lie within 30 pixels of the inner edges for the WFC chips and
within 50 pixels of the edge for the PC chip
\item exhibit a inverse flatfield value higher than 1.7, indicating
that nearly half of the total flux is lost because of proximity to the
pyramid edge
\item are adjacent to a pixel marked as saturated in the data quality file
 \end{enumerate}

For such bad pixels, we set the weight to zero. For the remaining
pixels, the weight is computed as the inverse of the variance,
according to the method of Casertano et al. (2000). This computation
of the variance takes into account contributions to the noise from the
sky background, dark current, read noise and flatfield. Contributions
to shot noise from sources are not included. These weight images are
used as the input weights for drizzle.

\subsubsection{First drizzle pass}

For each input image, a separate four-chip mosaic is generated using
the drizzle algorithm on an image that has the same frame as the
predefined output image.  At this stage, the image position and
orientation is defined using the header values for Chip 2; the
positions of the other three chips are adjusted according to the
time-dependent chip separation correction described in
Section~\ref{sect:alignment}.  Each image thus retains the
imperfections (cosmic rays, interchip gaps, etc.) of the input images,
and the registration between the images is based solely on the header
parameters.

\begin{figure*}
\plotone{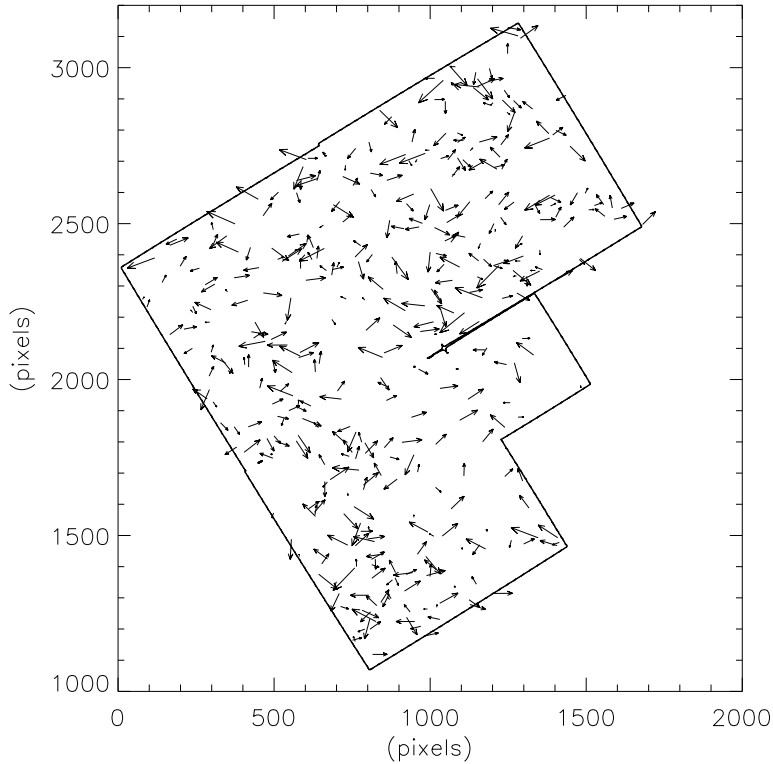} 
\caption{The vectors plotted show residual offsets in source positions between a reference image and a second image which has its coordinates transformed (with a small shift and rotation) to align it with the reference image. The offset vectors have been scaled up by a factor of 800 to make them visible. The outline is the WFPC2 footprint. It is clear that residuals are random and small (r.m.s.=7 milliarcsec). There are no chip-to-chip variations and no systematic variation within each chip.}\label{fig:smallshifts}
\end{figure*}

\begin{figure*}
\plotone{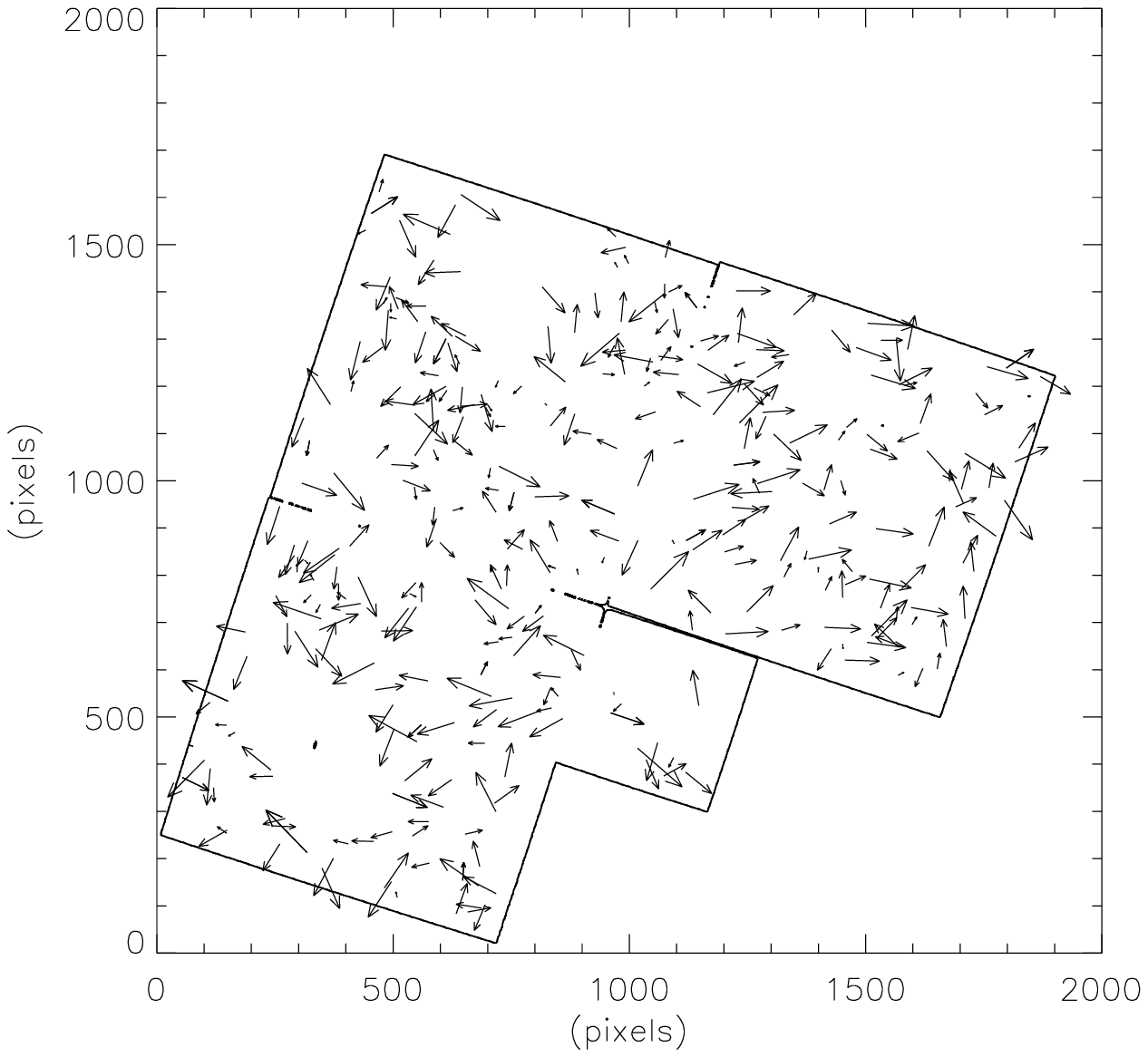} 
\caption{Vector plot of offsets in source positions between a reference image and the shifted/rotated image where the shifts are large ($\sim 3$ pixels in each coordinate). Vectors are scaled up by a factor of 800 as in \fig{smallshifts}. The deviations are still random. However, their amplitude is somewhat larger.}\label{fig:bigshifts}
\end{figure*}

\subsubsection{Intrafilter image registration}

In order to improve the registration between images taken in the same
filter, sources are identified using  SExtractor in the single
mosaiced images obtained form each dataset, using the weights obtained
in the first pass combination.  The algorithms described in
Section~\ref{sect:alignment} are then used to determine relative
shifts and rotations needed to optimize the relative registration
among images taken in the same filter. The alignment quality is
assessed using the r.m.s. scatter of the position residuals after outlier
rejection. We show in Figures~5 and 6, typical vector plots of
residual offsets between the reference image and a second image whose
coordinates have been transformed using the best-fit transformation
provided by {\it geomap}. The residuals are random and small. There
are no systematic chip-to-chip variations which would indicate an
error in the relative positioning of the 4 chips. There are no
systematic variations within each chip either which would indicate
errors in the distortion correction. For images that have a large
number of sources --such as those shown in the figures-- the r.m.s. of
the residual offset is typically 10 milliarcsec or smaller in each
coordinate. 

The improved alignment information is propagated back to the input
images, and their WCS parameters are updated accordingly.  The shift
and rotation required for each image and the r.m.s. of the residuals are
recorded in the log file.

\subsubsection{Second drizzle pass and sky equalization}

Once the images have been correctly registered to each other, the images
are drizzled through for the second time, again with all 4 chips of a
dataset drizzled to the output frame.  The sky level in each image is
adjusted to ensure that the relative backgrounds match, using the sky
background equalization technique described in
Section~\ref{sect:backequal}.  The goal of this step is not to zero out
the sky, which would lead to incorrect results whenever diffuse emission
is present, but simply to remove time-dependent background variations
which can adversely affect the quality of the combined images.  An
overall sky background consistent with one of the input images is
retained.  As a result, all images produced by our pipeline include a
contribution from the sky background as observed. 

\subsubsection{The median combined image and cosmic ray rejection}

The single, aligned, sky-equalized images obtained as a result of the
second drizzle pass are median-combined to remove the impact of cosmic
rays.  The median combined image is blotted back to each input image,
generating a reference image for each detector.  This reference image
and the original image data are compared, taking into account the
possibility of a net offset due to sky equalization.  Cosmic rays are
identified as described in Section~\ref{sect:crrej}, and the
corresponding pixels recorded in the cosmic ray masks.

\subsubsection{Third drizzle pass, first image output}

Once the cosmic ray rejection is complete, the third pass drizzle pass
is performed, using the same inputs as in the second pass, but with
the input weights modified to zero the weight of pixels identified as
cosmic rays.  Unlike the previous drizzle passes, all input images for
each filter are now drizzled onto a common output image.  This image,
which is a weighted combination of all data for the relevant filter,
is written to disk with a header that records the basic information of
the images that have been included in the combination.  Besides the
combined (science) image and the corresponding weight image, the input
weight file to the third drizzle pass is also recorded; this file will
be needed as input for the Stage 2 pipeline.  The updated WCS values
for each input image, after image registration, are written as a FITS
extension table of the science image.

In addition, many intermediate files---such as the median combined
image, the blotted medians, and the cosmic ray flags---can be written to
disk if the debug flag is set. 

\subsection{Stage 2 Pipeline}

Once all filters in a particular pointing have been processed through
the Stage 1 pipeline, a science image and weight image for each filter
are obtained.  However, these images may not be properly aligned across
filters, since the registration process is carried out independently for
each filter.  Cross-filter registration is performed by the Stage 2
pipeline. 

\subsubsection{Cross filter registration}

In the current pipeline implementation, we treat the F606W science
image to be the reference image for registration across filters.  All
final science images are registered to the F606W image using the
technique described in Section~\ref{sect:alignment}.  We determine a
list of sources for each filter using  SExtractor with the weight file
from the Stage 1 pipeline, determine the relative shifts and rotations
needed to align each filter with the F606W image, and propagate these
shifts and rotations back to the original images, updating the WCS
information separately for each chip.  Note that absolute astrometry
of the resulting image will still suffer from any inaccuracies present
in the original images, which are derived from guide star position
information through the knowledge of the HST focal plane solution at
that time.  In principle, the absolute astrometry could be improved by
matching sources in our images with those listed in the the Guide Star Catalog II; this step has not been implemented yet.

\subsubsection{Final drizzle}

When all image headers have been updated, the final {\it wdrizzle}
pass is performed for all filters except F606W, using the updated
alignment information.  A new drizzle pass is preferable to shifting
the previously obtained image because it introduces less image
degradation and noise correlation; since the final pass adopts the
weight information previously determined, including the identification
of cosmic ray events, it can be carried out efficiently with a modest
impact on computing resources.

The images resulting from this final pass, together with updated header
information and a table extension containing the updated WCS parameters
for all input images, are the versions that are made available to the 
community through the MAST site (see Section~\ref{subsect:delivery}).

\section{Quality verification and data delivery}
\subsection{Quality control}

Each image went through an extensive set of automated and human eye
based checks before delivery to the archive. Automated checks included
numerous flags for poor astrometric registrations (too few sources,
extremely large image shifts, high r.m.s. residuals after {\it geomap}
transformations etc.), anomalous background offsets and over-rejection
of cosmic rays. These automated checks were supplemented by human
checks.  A typical quality control process for a pointing would
include the following human check procedure:

\begin{enumerate}

\item Display the final science images in each filter. Verify that
   each science images is of acceptable quality. Any image anomalies
   such as bias jump, OTA Earth Reflection Pattern, PC1 Stray Light
   Patterns etc. (Biretta, Ritchie \& Rudloff 1995) seen in the
   science image are noted in a log file.

\item Register and blink through the final science images in each
filter to verify accurate cross-filter registration.

\item Display each weight image and
   examine for data quality. Issues to look for: excessive coincidence
   between bright sources and low weight (may indicate rejection of
   source cores); large areas of low weight not coinciding with image
   overlap (may indicate background problems, such as Earth cross pattern);
   excessively high or low values; gridding (should not occur with 
   {\tt pixfrac=1}).

\item If the science or weight images show some anomaly: e.g. incorrect
   CR rejection, poor registration etc., look at the skycube image 
   to diagnose the problem. Diagnosis or suspicion is noted in the log
   file for subsequent analysis.

\item Examine the log files for the registration process to confirm
that all images in the alignment sequence were registered with low r.m.s
residuals in the {\it geomap} alignment process.

\end{enumerate}

We re-examine images that fail the quality control processes. If the
issue is fixable with reprocessing, this is done. For approximately
5\% of the final images, quality issues remain unresolved; these
images are not delivered to the archive. The most common cause of
failure in quality control is when most (all) datasets contributing to
the final drizzled image show the Earth Cross pattern.

A significant contributor to the remaining uncertainty in astrometric
registration is the random error in source centroid positions computed
by  SExtractor. The most recent version of that software (version
2.4.3) claims to greatly reduce such errors by allowing for centroid
computation using a window function. The  SExtractor documentation
claims that the accuracy offered by the window function based
computation is comparable to that obtained from PSF fitting
software. We are currently in the process of testing the new version
of  SExtractor and its centroiding algorithm. If found appropriate, we
will use to it to replace the version of the software currently in use 
in our
image registration procedure.

\begin{figure*}
\plotone{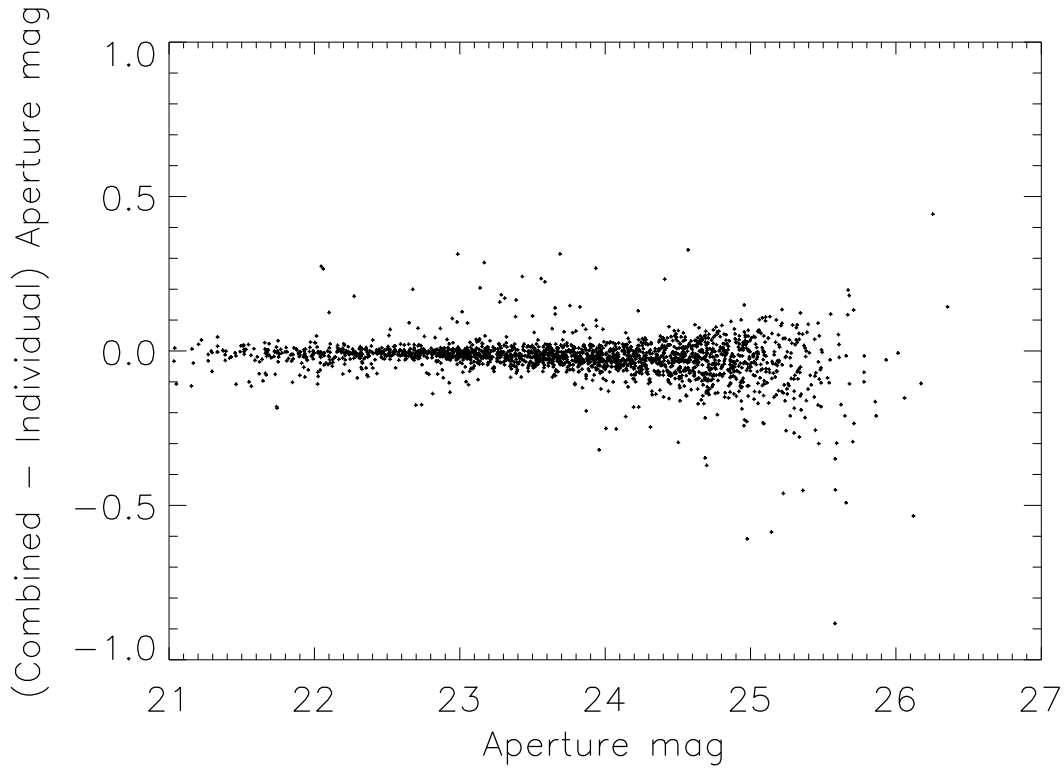} 
\caption{The difference in aperture magnitudes of sources in the combined image and a clipped mean of the same sources measured on the individual datasets that contributed to the combined image is plotted against the aperture magnitude for an image in the LMC. Sources for which aperture magnitudes are brighter in the individual images are almost certainly affected by cosmic rays. The remaining sources are distributed tightly about zero magnitude, with increased scatter at fainter magnitudes, as expected.}\label{fig:phototest}
\end{figure*}

\subsection{Photometry tests}

We tested a small number of images produced by our pipeline for
photometric accuracy. Our test consisted of comparing the instrumental
magnitudes of objects in the combined image with individual images
that contributed to it. The source detection and measurement
parameters in SExtractor were mostly set to default values, e.g.,
3-sigma thresholds for detection. Varying these parameters somewhat
did not change the basic result of the test. We used SExtractor in
dual image mode for the photometry. We detected objects in the
combined image and aperture magnitudes were measured in the combined
image and in each of the individual images obtained after the second
drizzle operation. We show in \fig{phototest} the plot of the
difference in aperture magnitude as measured on the combined image and
a clipped mean of the individual frames, as a function of aperture
magnitude in the combined image. A circular aperture with radius of 4
pixels was used.  The relative brightness of some objects in the
individual images is almost certainly due to cosmic ray contamination,
since the combined images are largely free of cosmic rays while the
individual images are not.

For all the images we tested, we found no systematic offsets in
photometry between the combined and individual images that would
indicate a problem in our procedure. We believe the combined images are
photometrically accurate and suitable for scientific
analysis. However, it should be noted that we have tested only a very
small fraction of the hundreds of images that we have produced.
Photometric errors in untested images may be present and may indicate
problems with our procedure; users discovering any such errors are urged
to contact us.

\subsection{The WFPC2 associations}

The Canadian Astronomy Data Center (CADC), the Space Telescope
European Coordination Facility (ST-ECF) and the Multimission Archive
at STScI (MAST) have made available a large number of combined WFPC2
images. These combined images are the products of the basic
registration and averaging of related sets of WFPC2 images, referred
to as associations. As of November 2002, over 15,000 combined images
had been created from associations of nearly 50,000 individual WFPC2
images.

The WFPC2 associations are a much stricter grouping of datasets
compared to the {\it pointings} in our project. Two (or more)
exposures in a given filter are grouped into an association if they
belong to the same program (same proposal id), their sky projected
distance is not greater than 10 arcsec (100 WF4 pixels) and their
position angle does not differ by more than 0.03 degrees. The APPP
places no restrictions on proposal id and position angle. The
separation in sky projected distance needs to be less than 80 arcsec
for the APPP. With our considerably looser criteria for grouping
images, in general, more datasets are grouped together in a
pointing. This implies a higher S/N in the final image. Also, by our
definition, any position on the sky produces only a single image in
each filter. For the WFPC2 associations, that is not necessarily true.

There are also significant differences in the data processing approach
taken by the two teams. The processing for the WFPC2 associations 
does not include drizzling
of the images or accounting for shifts in chip positions with
time. They also do not include distortion corrections. Their
procedures for image registration and cosmic-ray rejection are also
different. With our more ambitious approach, there are more avenues
for failure during the image combination process. To compensate for
this, we have included human eye checks as an integral part of our
quality control procedures.

Given the differences in grouping datasets together and the data
processing procedures between the two projects, a comparison of the
final images produced by the two projects is not
straightforward. Nevertheless, we have tried to compare our images
with those from the WFPC2 associations for a few representative pointings. We
find that both the images have S/N that is consistent with the
detector and sky background characteristics and the effective exposure
time.

Comparing the PSF's of starlike sources, we find that the FWHM of
sources in our images is not systematically different from the
corresponding sources in the WFPC2 associations. This is particularly
encouraging, because in principle our PSF may be broadened relative to
the WFPC2 associations by a combination of factors: resampling during
drizzling, errors in interchip registration, and in source-centroiding
in our image registration procedure.

\subsection{Data delivery and Web access}\label{subsect:delivery}

We deliver final science images, weight images and a log file
containing a brief summary of the properties of the images being
drizzled and their registration, as High Level Science Products (HLSP)
to the MAST science archive. In addition, we also provide a feature
rich, web based, front end to the data for easy browsing. This
includes a DSS image of an 18x18 arcmin area centered on the pointing,
a three color composite WFPC2 image made using F450W/F606W/F814W data
in the Blue/Green/Red channels respectively, header information for
each image and a preview image in each filter. For each pointing, we
also provide a coverage map in the 4 principal broadband filters. We
provide a link for each pointing to the NASA Extragalactic Database
(NED) which searches it for objects within 10 arcmin of the pointing
center. No source catalogs have been released yet, although we aim to
add them in future. Delivered data are grouped together by the science
questions that they are useful in addressing. Metadata from our
delivered images have been incorporated into the MAST database and is
searchable through the the MAST search interface. Image header
keywords (and other metadata) will be updated on an ongoing basis to
make the data more accessible through the Virtual Observatory. Any
modifications to the procedures described in this paper, will be fully
documented in the README files accompanying data released through
the MAST archive.

\acknowledgements

We are grateful to the anonymous referee whose insightful comments
greatly improved this paper. Support for program AR 9540 was provided
by NASA through a grant from the Space Telescope Science Institute,
which is operated by the Association of Universities for Research in
Astronomy, Inc., under NASA contract NAS 5-26555.

\end{document}